\documentclass[12pt]{article}
\usepackage[margin=1in]{geometry}
\usepackage{natbib}
\usepackage{setspace}
\usepackage[skip=10pt plus1pt, indent=20pt]{parskip}

\PassOptionsToPackage{hyphens}{url}
\usepackage[
  colorlinks=true,
  linkcolor=black,
  citecolor=blue,
  urlcolor=blue,
  breaklinks=true,
]{hyperref}
\usepackage{xurl}
\usepackage{adjustbox}
\usepackage{caption}
\usepackage{tabularx}
\usepackage{amsmath}
\usepackage{amssymb}
\usepackage{lmodern}
\usepackage{microtype}
\usepackage{longtable}
\usepackage{array}
\usepackage{booktabs}
\usepackage{graphicx}
\usepackage{float}
\graphicspath{{figs/}}

\title{The Cognitive Kardashev Scale: \\
        \Large Quantifying the Material Envelope of Civilisational Computation}
\author{Sachin Sharma\thanks{\href{mailto:sachin.econix@gmail.com}{\texttt{sachin.econix@gmail.com}}}}
\date{}

\begin{document}

\maketitle

\begin{abstract}
\noindent How much thinking can a civilisation do? Kardashev ranked civilisations by the energy they command. This paper borrows his ladder and asks how much machine cognition each rung could support. The arithmetic is deliberately simple. A civilisation has some total power. Only a fraction of that can be spared for computing, and each joule spent buys computation at whatever efficiency the hardware of the day has reached. The product of the three sets a ceiling on machine thought. To keep the resulting quantities intelligible, I express them in units of the human brain's own processing rate, as a rough yardstick rather than a claim about minds. Calibrating the ceiling against present-day supercomputers and AI accelerators led me to two conclusions I did not expect at the outset. Even today's energy supply could support far more machine cognition than humanity actually uses, so physical capacity is not what binds. And whether energy or hardware efficiency becomes the constraint over the coming decade turns on engineering choices that have not yet been made. On the question that may matter most, who gets access to the cognition it describes, the scale is silent. That is a matter of political economy, and the calibration is offered as an input to that debate.
\end{abstract}

\section{Introduction}

Humanity already lives inside an extended cognitive system whose energy footprint rivals that of medium-sized countries. Global data centres consumed $\sim$415 TWh of electricity in 2024 and are projected to roughly double by 2030 \citep{IEA_EnergyAI_2025}. A single frontier AI training run now draws power on the order of a small city for weeks at a time. Most of this compute runs on NVIDIA's Vera Rubin platform, which enters volume production in 2026 at $\sim$50 PFLOP/s sparse FP4 per accelerator, and it is deployed inside vertically integrated, hundred-billion-dollar consortia whose announced infrastructure commitments (Stargate, Terafab) exceed any previous industrial buildout \citep{Stargate_2025,BloombergTerafab2026}. These investments are already under way. The open question, and the one this paper asks, is what cognitive capacity they could in principle support and how that capacity scales as energy budgets grow.

\citet{kardashev1964} proposed that civilisations be classified by total power consumption. Type~I commands the energy striking a planet from its parent star ($\sim 10^{16}$ W); Type~II commands the full luminosity of its star ($\sim 10^{26}$ W); Type~III commands a galaxy ($\sim 10^{37}$ W). Kardashev framed the scale as a tool for SETI, oriented toward whether a civilisation's energy output would be detectable across interstellar distances. Carl Sagan later refined the scale to a continuous logarithmic index $K = (\log_{10} P - 6)/10$, which places contemporary humanity, with primary energy consumption $\sim 2\times10^{13}$ W in 2024 \citep{IEA_WEO_2024}, at $K \approx 0.73$. By Sagan's measure, humanity has travelled roughly three-quarters of the way from a pre-industrial baseline ($K = 0$) to commanding the full energy budget of a planet ($K = 1$).

The original Kardashev scale remains an energy and detectability typology, and this paper does not modify it. I instead construct an analogous framework, which I call the \textit{Cognitive Kardashev Scale}, that runs in parallel with Kardashev's original tiers but is calibrated in computational units rather than purely energetic ones. The contribution is calibration rather than new theory. The paper reports what total sustained machine-learning-grade compute, in floating-point operations per second, each Kardashev tier could support given a defensible 2024--2026 efficiency anchor. The arithmetic is straightforward. Take the total power $P$ a civilisation has available (measured in watts, i.e.\ joules per second). A civilisation does not allocate \textit{all} of that to thinking; it also has to grow food, run factories, and move people. Let $f$ be the fraction set aside for cognition. Computation costs energy, at an efficiency that improves over time; let $\eta$ be the rate at which a system converts joules of energy into operations of compute (operations per joule). Then the available cognitive throughput is
\begin{equation}\label{eq:basic}
C_{\text{cog}} \;=\; f\,P\,\eta \quad \text{(operations per second)}.
\end{equation}
This is the primary quantity reported throughout the paper. To make the result intuitive rather than just a large number, I also express $C_{\text{cog}}$ in human-brain-equivalent units, dividing by $C_{\text{brain}}$ (the brain's own processing rate, in operations per second). The secondary quantity, $N = C_{\text{cog}}/C_{\text{brain}}$, is reported with explicit uncertainty, because the literature on $C_{\text{brain}}$ spans $10^{15}$--$10^{17}$ FLOP/s depending on which neural events are counted as ``operations,'' so brain-equivalent counts carry at minimum $\pm 1$ order of magnitude of irreducible uncertainty. I use them as a yardstick, not as a literal claim that so many FLOP/s of silicon reproduces a human mind. The cognitive-Kardashev framework and the original Kardashev scale are therefore complementary; the former indexes how much sustained ML-grade compute a given Kardashev-tier energy budget can in principle support, holding the latter's energetic typology fixed.

The paper is organised as follows. Section~\ref{sec:eta} anchors the efficiency parameter $\eta$ in 2024--2026 hardware data, comparing six classes of workload from sustained scientific FP64 supercomputing to the Landauer thermodynamic limit. Section~\ref{sec:brain} adopts a central biological-baseline estimate. Section~\ref{sec:earth} establishes the Earth 2024--25 baseline. Section~\ref{sec:envelope} computes the Type~I, II, and III cognitive envelopes and presents the scale. Section~\ref{sec:fractional} reports per-capita capacity under realistic fractional allocation. Section~\ref{sec:trajectory} presents three forward scenarios for frontier training compute and connects them to current infrastructure commitments. Section~\ref{sec:sensitivity} discusses parameter sensitivity. Section~\ref{sec:hayek} discusses what the Scale does and does not show about the political economy of cognitive compute, and Section~\ref{sec:conclusion} concludes.

\section{Compute Efficiency in 2026}\label{sec:eta}

\begin{figure}[!htbp]
\centering
\includegraphics[width=0.95\textwidth]{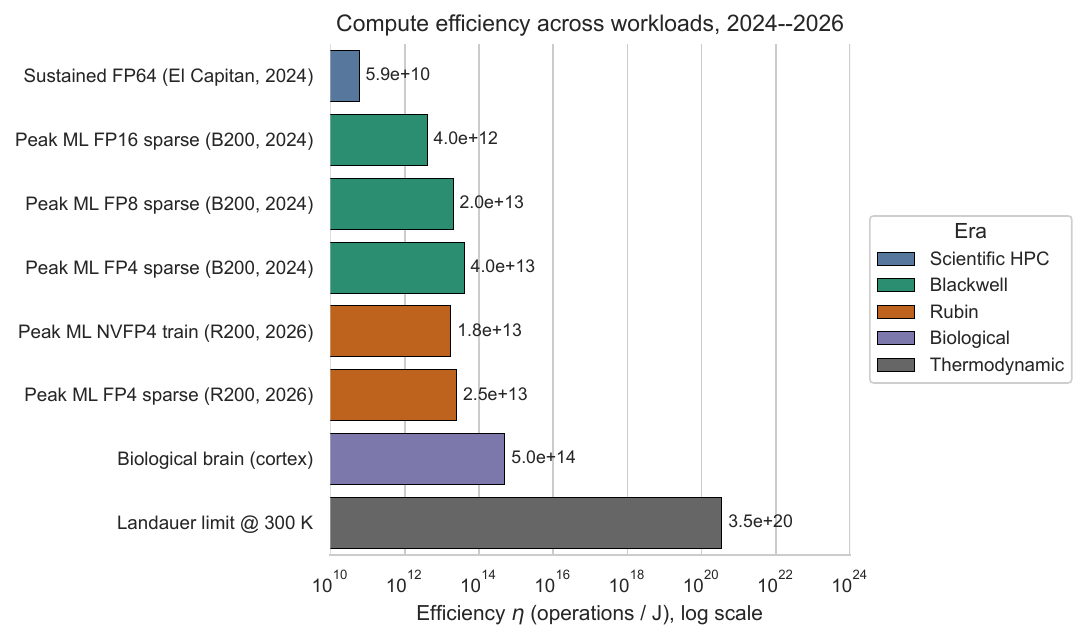}
\caption{\textbf{Compute efficiency $\eta$ across workload classes, 2024--2026.} Bars on a logarithmic horizontal axis. The roughly two- to three-order-of-magnitude gap between sustained scientific (FP64) and peak ML inference (FP4) reflects the move to lower-precision arithmetic; the further gap of up to two orders of magnitude between peak ML inference and the upper end of biological-brain efficiency bounds the energetic advantage that biology retains over silicon at the single-system level; and the approximately six-order-of-magnitude gap between the brain and the Landauer limit at 300~K records the headroom available to thermodynamically optimal computation. None of the values in Tables~\ref{tab:envelope}--\ref{tab:big} approach that bound.}
\label{fig:eta}
\end{figure}

The relevant efficiencies, expressed as floating-point operations per joule of electrical input, span roughly twelve orders of magnitude from sustained scientific computing through near-term ML accelerators to the Landauer thermodynamic limit (Table~\ref{tab:eta}; Figure~\ref{fig:eta}).

\begin{table}[H]
\centering
\caption{Compute efficiency $\eta$ in 2026, by class of workload. ``Sustained'' denotes top-of-list HPL benchmarks; ``peak ML'' denotes vendor-reported sparse Tensor Core throughput at design power.}\label{tab:eta}
\small
\begin{tabular}{lcc}
\toprule
Workload class & Representative system & $\eta$ (operations / J) \\
\midrule
Sustained scientific (FP64) & El~Capitan, LLNL (2024) & $5.9\times10^{10}$ \\
Peak ML training (FP16) & NVIDIA B200 sparse (2024) & $\sim 4\times10^{12}$ \\
Peak ML inference (FP8) & NVIDIA B200 sparse (2024) & $\sim 2\times10^{13}$ \\
Peak ML inference (FP4) & NVIDIA B200 sparse (2024) & $\sim 4\times10^{13}$ \\
Peak ML inference (FP4) & NVIDIA R200 sparse (2026) & $\sim 2.5\times10^{13}$ \\
Peak ML training (NVFP4) & NVIDIA R200 sparse (2026) & $\sim 1.8\times10^{13}$ \\
Biological brain & Human cortex ($\sim$20 W) & $5\times10^{13}$--$5\times10^{15}$ \\
Landauer limit (300 K) & Irreversible bit-op bound & $\sim 3.5\times10^{20}$ \\
\bottomrule
\end{tabular}
\end{table}

The Landauer entry is in irreversible bit-operations per joule, given by $1/(k_{\mathrm B} T \ln 2) = 3.48\times10^{20}$ at $T=300$~K \citep{Landauer1961}. The other rows are in floating-point operations per joule. Because a single double- or half-precision FLOP corresponds to many bit-operations of internal state-flipping, the Landauer-limit FLOP/J is much smaller than the bit-operations-per-joule figure quoted; the comparison across rows is therefore qualitative, not strictly commensurable. \citet{Lloyd2000} establishes higher physical bounds on information processing in mass--energy systems that lie further still beyond the values in Table~\ref{tab:eta}.

Two notes on the Blackwell-to-Rubin transition are needed. First, Rubin's per-chip FLOP/J figures are not strictly higher than Blackwell's, because TDP rose from $\sim$1\,kW (B200) to $\sim$1.8--2.3\,kW (R200) per accelerator. Rubin's claimed efficiency advantage is realised at the rack and inference-throughput level, where much larger HBM4 memory bandwidth ($\sim$22 TB/s vs $\sim$8 TB/s on B200) and tighter NVLink fabric reduce data-movement overhead substantially \citep{NVIDIA_Rubin_2026}. Second, the hierarchy in Table~\ref{tab:eta} is internally consistent across workload precision; each step from FP64 down to FP4 doubles to quadruples efficiency, broadly matching vendor claims of $2$--$4\times$ per precision halving once memory and utilisation overheads are properly accounted.

For the cognitive recalculation in this paper, the relevant efficiency is the rate at which energy can be converted into machine-learning-grade compute, since the kind of cognition under discussion is overwhelmingly performed at FP16 or lower precision in current and near-future systems. I adopt
\[
\eta_{2026} \;=\; 1\times 10^{12}\ \text{FLOP/J}
\]
as a defensible 2026 anchor. This sits roughly an order of magnitude below B200 peak sparse FP16 and an order of magnitude below R200 peak sparse FP8, because real workloads include data movement, memory access, network traffic, cooling overhead, and utilisation losses that bring sustained efficiency well below vendor peaks \citep{Strubell2019}. It is also broadly consistent with reported tokens-per-joule efficiencies for production-scale inference on Blackwell- and Rubin-class systems, in the range $10^{12}$--$10^{13}$ effective FLOP/J once memory and fabric overhead is included.

The biological brain remains, on this measure, between roughly $1.5$ and $3.5$ orders of magnitude more efficient per joule than current sustained digital ML substrates. To make that gap concrete, a human cortex running on $\sim$20 watts, about the power draw of a dim incandescent bulb, outperforms several hundred to several thousand watts of contemporary AI hardware on the same kind of pattern-recognition task. A decade of rapid hardware progress has not changed this ranking. This margin matters for two reasons. First, it bounds how much of the Kardashev energy budget any digital cognitive architecture can in principle convert into useful thought relative to a hypothetical biological substrate. Second, it means that ``efficiency progress'' over the next several decades has a clear destination, the silicon-to-cortex gap. Whether engineering gets there, or stalls well short, is the open empirical question that determines whether the binding constraint on civilisational cognition is energy or efficiency. Sensitivity to $\eta$ is one-to-one and explored in Section~\ref{sec:sensitivity}.

\section{Brain Reference Values}\label{sec:brain}

Synaptic-count and molecular-state analyses attribute approximately $10^{14}$--$10^{15}$ synapses to the human cortex, with roughly 4.7 distinguishable states per synapse \citep{Bartol2015,Drachman2005}. Adopting an order-of-magnitude estimate,
\[
M_{\text{brain}} \;\approx\; 10^{15}\ \text{bytes}.
\]
For processing capacity, estimates vary from $10^{15}$ to $10^{17}$ operations per second depending on the level at which one counts neural events as ``operations'' \citep{LevyCalvert2021,SandbergBostrom2008}. The literature uses ``operations'' loosely; mappings to floating-point operations are inevitably approximate. I adopt a central value of
\[
C_{\text{brain}} \;=\; 10^{16}\ \text{FLOP/s},
\]
in line with computational-neuroscience reviews that count synaptic events as floating-point operations and adjust for the fact that $\sim$95\% of neural energy is spent on communication rather than the computational equivalent of multiply-accumulate \citep{LevyCalvert2021}. Brain-equivalent counts in this paper should be read as carrying $\pm 1$ order of magnitude of irreducible uncertainty from this source; full propagation across all parameters is reported in Section~\ref{sec:sensitivity}. The convention adopted is that ``brain-equivalents'' is an interpretable reference unit, not a literal claim that $C_{\text{brain}}$ FLOP/s of digital compute reproduces a human mind. Whole-brain emulation \citep{SandbergBostrom2008} would require structural and data conditions not addressed here.\footnote{For a memory-side comparator, the global datasphere reached $\sim 1.75$--$2.0\times 10^{23}$ bytes in 2024 against $M_{\text{brain}} \approx 10^{15}$ bytes per cortex, a ratio of $\sim 10^{8}$. I do not use this comparison further; the cognitive claims in this paper concern processing throughput, not storage.}

The hundred-fold spread in $C_{\text{brain}}$ deserves comment. The literature spans $10^{15}$ to $10^{17}$ FLOP/s because the question ``how much computation does a brain do?'' has no canonical answer. Counting only the spike events that cross synapses gives the low end of the range; adding the dendritic integration upstream of those spikes, and then the slower neuromodulatory and glial dynamics, moves the estimate up by orders of magnitude. A two-orders-of-magnitude range over what a single cortex represents is a measure of how much remains unknown about cognition. The brain-equivalent is therefore a deliberately rough yardstick, a way of asking whether a given number of FLOPs corresponds to something like one mind or a million, rather than a precise translation. Throughout the rest of the paper I use the central value and report the band; readers should treat conclusions about specific brain-equivalent counts as good to within a factor of ten, and conclusions about the qualitative ordering of Kardashev tiers as robust.

\section{Earth 2024--25 Baseline (Type 0.73)}\label{sec:earth}

Global primary energy consumption in 2024 was approximately $620$ EJ/yr, equivalent to $\sim 2.0\times10^{13}$ W of continuous power \citep{IEA_WEO_2024}. On Sagan's logarithmic refinement of the Kardashev scale ($K = (\log_{10} P - 6)/10$), this places contemporary humanity at $K \approx 0.73$. Of this, data centres accounted for $\sim$415 TWh in 2024, or about $4.7\times10^{10}$ W continuous, roughly $0.24\%$ of total primary energy \citep{IEA_EnergyAI_2025}. The IEA's Base Case projects this rising to $\sim$945 TWh ($\sim 1.1\times10^{11}$ W) by 2030 \citep{IEA_EnergyAI_2025,IEA_DataCenterSurge2025}.

Two intermediate adjustments are needed before applying $\eta_{2026}$. First, the data-centre electricity figure includes facility overhead (cooling, power conversion, network) at a typical Power Usage Effectiveness (PUE) of $\sim 1.2$--$1.5$, meaning $20\%$--$33\%$ of the input is consumed by non-compute infrastructure. Compute-effective power is therefore $P_{\text{compute}} = P_{\text{DC}} / \mathrm{PUE}$. Second, only a fraction of compute-effective workloads are ML-grade in any meaningful sense; cloud serving, video streaming, and conventional enterprise workloads dominate at present, with AI accelerators growing rapidly but still a minority share of installed compute. I denote this fraction $\phi$ and treat it as a free parameter; for the global aggregate I take $\phi \in [0.05, 0.30]$ as a plausible 2024 range, with the upper bound rising as AI buildout proceeds.

Applying $\eta_{2026} = 10^{12}$ FLOP/J to the compute-effective, ML-grade share gives
\[
C_{\text{ML,2024}} \;\approx\; \frac{\phi\,P_{\text{DC}}}{\mathrm{PUE}}\,\eta_{2026} \;=\; \frac{\phi \cdot 4.7\times 10^{10}\,\text{W}}{1.3} \cdot 10^{12}\,\text{FLOP/J}.
\]
For $\phi \in [0.05, 0.30]$ and $\mathrm{PUE} = 1.3$, this yields $C_{\text{ML,2024}} \in [1.8, 11]\times 10^{21}$ FLOP/s, equivalent to $\sim 10^{5}$--$10^{7}$ brain-equivalents under the $C_{\text{brain}}$ uncertainty range. The wide bracket reflects genuine uncertainty in the AI-share of installed compute; it is not a defect of the analysis but a feature of the underlying empirical situation. By 2030, on the IEA Base Case projection of data-centre electricity to $\sim 945$ TWh ($\sim 1.1\times 10^{11}$ W) and an upward revision of $\phi$ toward the high end of its range, the equivalent figure rises to roughly $10^{7}$--$10^{8}$ brain-equivalents in raw capacity. These figures should not be interpreted as a claim that the existing digital infrastructure already constitutes that many ``minds''; they are physical-capacity ceilings, not realised cognitive throughput.

In human-scale terms, the midpoint of the 2024 estimate, $\sim 10^{6}$ brain-equivalents of ML-grade compute, is comparable to the population of a major metropolitan area. Most of this compute does nothing one would recognise as cognition; it serves video, processes payments, and runs enterprise software. But the share devoted to AI inference, training, and decision-support is rising rapidly. A frontier large-language-model service operating continuously at the scale of a few hundred megawatts, comparable to a single Stargate-class data centre, already commands cognitive capacity equivalent, in raw FLOP/s, to a small city's worth of human cortex. When one asks whether AI is becoming a cognitive utility on the scale of electrification or literacy, the order-of-magnitude arithmetic is already there. What remains contingent is who has access to it and on what terms.

\section{Type I, II, and III: Cognitive Capacity Envelopes}\label{sec:envelope}

\begin{figure}[!htbp]
\centering
\includegraphics[width=0.95\textwidth]{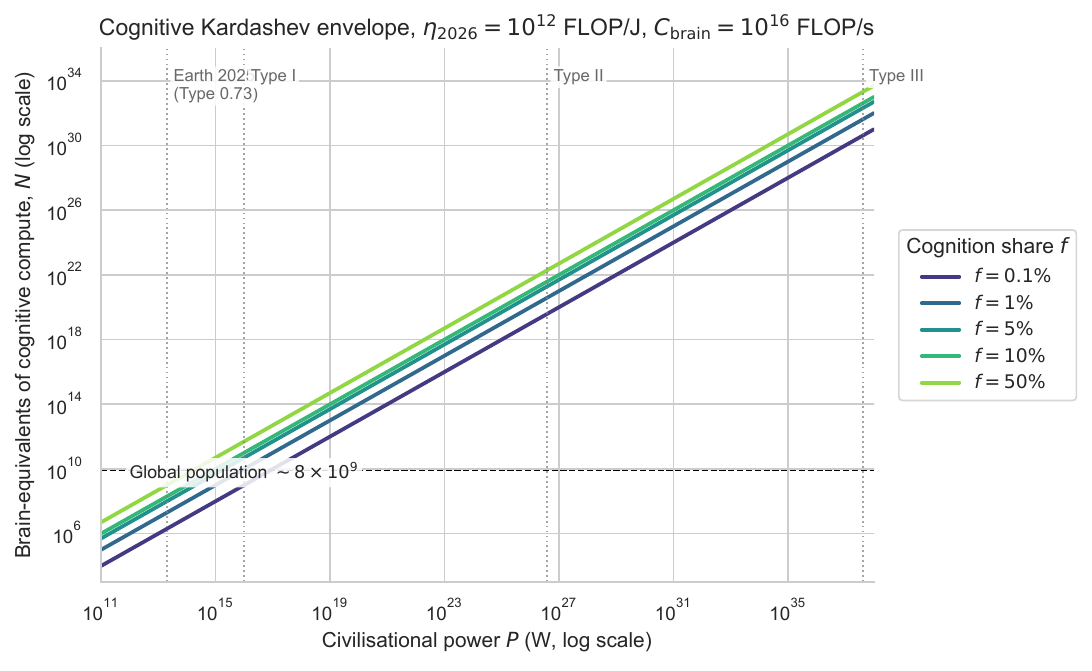}
\caption{\textbf{Cognitive Kardashev Scale.} Total brain-equivalents $N$ as a function of civilisational power $P$, at five energy-allocation fractions $f \in \{0.1\%, 1\%, 5\%, 10\%, 50\%\}$. Computed at $\eta=10^{12}$ FLOP/J and $C_{\text{brain}}=10^{16}$ FLOP/s; brain-equivalent values carry roughly $\pm 1$ order of magnitude of irreducible uncertainty from the $C_{\text{brain}}$ literature range. Vertical dotted lines mark the four reference power levels (Earth 2025, Type~I, Type~II, Type~III); the dashed horizontal line at $N = 8.1\times10^{9}$ marks current global population. Type~I at $f = 1\%$ already crosses parity with the human population at the central $C_{\text{brain}}$ value; Type~II at modest $f$ is a few billion brain-equivalents per human inhabitant.}
\label{fig:kardashev_curve}
\end{figure}

Recall the core arithmetic. The available cognitive throughput of a civilisation is the product of how much power it has, $P$ (in watts), and how efficiently it can convert that power into computation, $\eta$ (in operations per joule),
\[
C_{\text{cog}} \;=\; P\,\eta \quad \text{(operations per second)},
\]
and the corresponding brain-equivalent count is $N = C_{\text{cog}}/C_{\text{brain}}$, where $C_{\text{brain}}$ is the brain's own processing rate in the same units. Table~\ref{tab:envelope} reports the resulting envelope for Kardashev Types~I, II, and III at $\eta_{2026} = 10^{12}$ FLOP/J and $C_{\text{brain}} = 10^{16}$ FLOP/s; Figure~\ref{fig:kardashev_curve} plots the same data continuously across $P$ at five values of $f$.

\begin{table}[H]
\centering
\caption{Cognitive capacity at full energy allocation, Kardashev Types I--III, at central parameter values $\eta = 10^{12}$ FLOP/J and $C_{\text{brain}} = 10^{16}$ FLOP/s. The brain-equivalent column $N$ carries roughly $\pm 1$ order of magnitude of irreducible uncertainty from the literature range on $C_{\text{brain}}$ alone (see Section~\ref{sec:sensitivity}); the FLOP/s column $C_{\text{cog}}$ is correspondingly tighter.}\label{tab:envelope}
\small
\begin{tabular}{lccc}
\toprule
Civilisation & Power $P$ (W) & $C_{\text{cog}} = P\eta$ (FLOP/s) & Brain-equivalents $N$ \\
\midrule
Type 0.73 (Earth 2025) & $2.0\times10^{13}$ & $2.0\times10^{25}$ & $2.0\times10^{9}$ \\
Type I (planetary) & $1\times10^{16}$ & $1\times10^{28}$ & $1\times10^{12}$ \\
Type II (stellar) & $3.8\times10^{26}$ & $3.8\times10^{38}$ & $3.8\times10^{22}$ \\
Type III (galactic) & $4\times10^{37}$ & $4\times10^{49}$ & $4\times10^{33}$ \\
\bottomrule
\end{tabular}
\end{table}

Three observations follow. First, even at present terrestrial power, a hypothetical full allocation of energy to cognition would already support $\sim 2\times 10^{25}$ FLOP/s, equivalent to $\sim 10^{8}$--$10^{10}$ brain-equivalents under the $C_{\text{brain}}$ uncertainty range. The ratio of actual to maximum is overwhelmingly an allocation problem rather than a physical-capacity problem. Second, the gap from Earth 2025 to Type~I is a factor of $\sim$500 in cognitive capacity (between two and three orders of magnitude), the gap from Type~I to Type~II is ten orders of magnitude, and Type~II to Type~III another eleven. Each Kardashev step moves cognitive capacity into a qualitatively different regime. Third, the Landauer bound \citep{Landauer1961}, $1/(k_{\mathrm B} T \ln 2) \approx 3.5\times10^{20}$ irreversible bit-operations per joule at 300~K, lies many orders of magnitude above any value in Table~\ref{tab:envelope}. This does not mean current and projected systems enjoy a useful safety margin; it means real computation operates far less efficiently than thermodynamics would in principle permit. Whether engineering can close any meaningful fraction of the gap between sustained ML-grade compute and Landauer is a hardware-physics question well beyond this paper's scope; I report the bound only as a logical ceiling, not as a target.

The Kardashev tiers translate into very different human futures. A Type~I civilisation at $f = 1\%$ commands $\sim 10^{10}$ brain-equivalents of digital cognition (at the central reference value), roughly one digital mind per inhabitant, spread across personal assistants, scientific search, and routine decision-support. At $f = 10\%$ the figure is ten per inhabitant. Type~II lies so far beyond present experience that its numbers resist intuition. At $f = 1\%$, the per-inhabitant cognitive surplus is several billion brain-equivalents, comparable to the entire human population per individual citizen. Type~III is, on this measure, mostly a placeholder for ``what stellar engineering could do if it were possible.'' I include it for completeness, but draw no policy conclusions from it. The substantive lesson of the envelope is that humanity's accessible cognitive future, on any reasonable horizon, lies between Type 0.73 (where humanity stands today) and Type~I (where planetary solar capture would put it). Everything beyond that is a thought experiment.

\section{Realistic Per-Capita Cognition under Fractional Allocation}\label{sec:fractional}

\begin{figure}[!htbp]
\centering
\includegraphics[width=\textwidth]{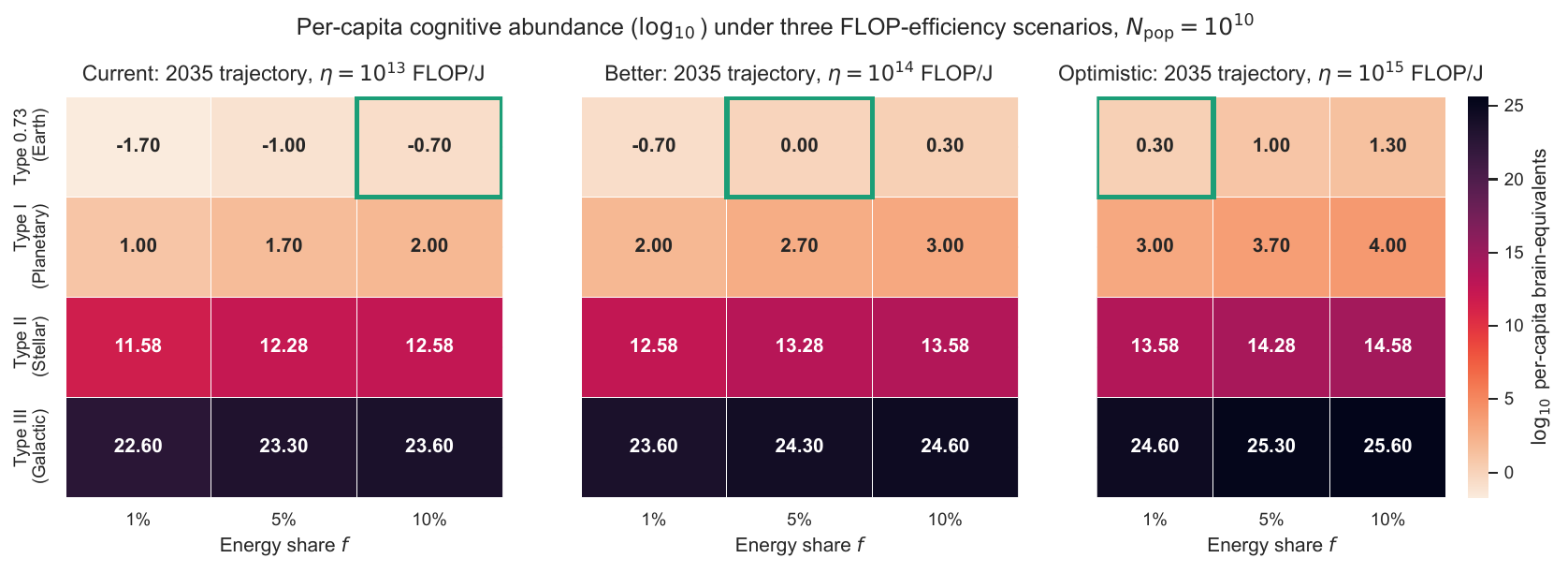}
\caption{\textbf{Per-capita cognitive abundance ($\log_{10}$) under the three Figure~\ref{fig:trajectory} scenarios projected to 2035.} Three panels share the same Kardashev-tier $\times$ allocation-share grid; the cell value is the logarithm (base 10) of the number of brain-equivalents per human inhabitant, assuming $N_{\text{pop}} = 10^{10}$. \textit{Current} (left, $\eta_{2035} = 10^{13}$ FLOP/J): the 2035 endpoint of the Figure~\ref{fig:trajectory} regression-based trajectory, attributing roughly 10$\times$ of the underlying nine-year compute growth to efficiency improvement and the rest to chip-count and run-duration scaling. \textit{Better} (centre, $\eta_{2035} = 10^{14}$ FLOP/J): the 2035 endpoint of the Figure~\ref{fig:trajectory} Sevilla-rate trajectory, with $\sim$100$\times$ efficiency improvement consistent with neuromorphic-adjacent ML hardware. \textit{Optimistic} (right, $\eta_{2035} = 10^{15}$ FLOP/J): the 2035 endpoint of the Figure~\ref{fig:trajectory} capacity-led trajectory, requiring $\sim$1000$\times$ efficiency improvement and reaching the lower bound of biological-cortex efficiency. Green borders highlight the cell within each panel closest to per-capita parity (one brain-equivalent per inhabitant). The parity threshold migrates leftward and downward as efficiency improves under the three Figure~\ref{fig:trajectory} trajectories: from a near-parity Earth-scale cell at $f = 10\%$ in \textit{Current}, to exact parity at Earth $\times$ $f = 5\%$ in \textit{Better}, to above-parity Earth at $f = 1\%$ in \textit{Optimistic}.}
\label{fig:heatmap}
\end{figure}

Real civilisations do not allocate all their energy to cognition. Agriculture, manufacturing, infrastructure, mobility, and the environmental and biospheric overheads of civilisation itself absorb the remainder.\footnote{Energy-allocation overheads at planetary scale are reviewed in the IEA's annual outlooks \citep{IEA_WEO_2024,IEA_EnergyAI_2025}; the 1\%--10\% range used here is a conservative bracket. The 2024 data-centre share of $\sim$0.24\% of global primary energy is well below 1\%, but compute is the most rapidly growing component of demand, plausibly reaching a few per cent by mid-century.} Let $f$ denote the fraction of total power allocated to cognition. Per-capita cognitive capacity, dividing the cognitive throughput across the population $N_{\text{pop}}$, is
\[
C_{\text{per capita}} \;=\; \frac{f\,P\,\eta}{N_{\text{pop}}}.
\]
For round-number arithmetic I assume a stable $N_{\text{pop}} = 10^{10}$ across the table and the heatmap (current world population is $8.1\times 10^{9}$ in 2024 and projected by the UN to $\sim 9.7\times 10^{9}$ by 2050, so the implied error is $\sim 25\%$, well below the irreducible uncertainty in $C_{\text{brain}}$ and $\phi$). Table~\ref{tab:big} reports total and per-capita brain-equivalents at the central parameter values, and the corresponding logarithmic cognitive Kardashev indices $\mathcal{K}^{\text{tot}}_{\text{cog}} = \log_{10}(N_{\text{tot}})$ and $\mathcal{K}^{\text{pc}}_{\text{cog}} = \log_{10}(N_{\text{pc}})$ at $f \in \{1\%, 5\%, 10\%\}$. Figure~\ref{fig:heatmap} visualises the full grid on a $\log_{10}$ heatmap. Both should be read with the $\pm 1$-order-of-magnitude band described in Section~\ref{sec:sensitivity}.

\begin{table}[H]
\centering
\scriptsize
\caption{Cognitive Kardashev Scale at $f \in \{1\%, 5\%, 10\%\}$, with $\eta = 10^{12}$ FLOP/J, $C_{\text{brain}} = 10^{16}$ FLOP/s, $N_{\text{pop}} = 10^{10}$. $\mathcal{K}^{\text{tot}}_{\text{cog}} = \log_{10}(N_{\text{tot}})$, $\mathcal{K}^{\text{pc}}_{\text{cog}} = \log_{10}(N_{\text{pc}})$.}\label{tab:big}
\begin{tabular}{lcccccccc}
\toprule
\textbf{Level} & $f$ & $P_{\text{tot}}$ (W) & $P_{\text{cog}}$ (W) & $C_{\text{cog}}$ (FLOP/s) & $N_{\text{tot}}$ & $N_{\text{pc}}$ & $\mathcal{K}^{\text{tot}}_{\text{cog}}$ & $\mathcal{K}^{\text{pc}}_{\text{cog}}$ \\
\midrule
\multicolumn{9}{c}{\textit{Type 0.73 (Earth 2024--25), $P_{\text{tot}} = 2\times10^{13}$ W}} \\
\midrule
Type 0.73 & 1\% & $2\times10^{13}$ & $2.0\times10^{11}$ & $2.0\times10^{23}$ & $2.0\times10^{7}$ & $2.0\times10^{-3}$ & 7.30 & $-2.70$ \\
Type 0.73 & 5\% & $2\times10^{13}$ & $1.0\times10^{12}$ & $1.0\times10^{24}$ & $1.0\times10^{8}$ & $1.0\times10^{-2}$ & 8.00 & $-2.00$ \\
Type 0.73 & 10\% & $2\times10^{13}$ & $2.0\times10^{12}$ & $2.0\times10^{24}$ & $2.0\times10^{8}$ & $2.0\times10^{-2}$ & 8.30 & $-1.70$ \\
\midrule
\multicolumn{9}{c}{\textit{Type I (planetary), $P_{\text{tot}} = 1\times10^{16}$ W}} \\
\midrule
Type I & 1\% & $1\times10^{16}$ & $1.0\times10^{14}$ & $1.0\times10^{26}$ & $1.0\times10^{10}$ & $1.0$ & 10.00 & 0.00 \\
Type I & 5\% & $1\times10^{16}$ & $5.0\times10^{14}$ & $5.0\times10^{26}$ & $5.0\times10^{10}$ & $5.0$ & 10.70 & 0.70 \\
Type I & 10\% & $1\times10^{16}$ & $1.0\times10^{15}$ & $1.0\times10^{27}$ & $1.0\times10^{11}$ & $10.0$ & 11.00 & 1.00 \\
\midrule
\multicolumn{9}{c}{\textit{Type II (stellar), $P_{\text{tot}} = 3.8\times10^{26}$ W}} \\
\midrule
Type II & 1\% & $3.8\times10^{26}$ & $3.8\times10^{24}$ & $3.8\times10^{36}$ & $3.8\times10^{20}$ & $3.8\times10^{10}$ & 20.58 & 10.58 \\
Type II & 5\% & $3.8\times10^{26}$ & $1.9\times10^{25}$ & $1.9\times10^{37}$ & $1.9\times10^{21}$ & $1.9\times10^{11}$ & 21.28 & 11.28 \\
Type II & 10\% & $3.8\times10^{26}$ & $3.8\times10^{25}$ & $3.8\times10^{37}$ & $3.8\times10^{21}$ & $3.8\times10^{11}$ & 21.58 & 11.58 \\
\midrule
\multicolumn{9}{c}{\textit{Type III (galactic), $P_{\text{tot}} = 4\times10^{37}$ W}} \\
\midrule
Type III & 1\% & $4\times10^{37}$ & $4.0\times10^{35}$ & $4.0\times10^{47}$ & $4.0\times10^{31}$ & $4.0\times10^{21}$ & 31.60 & 21.60 \\
Type III & 5\% & $4\times10^{37}$ & $2.0\times10^{36}$ & $2.0\times10^{48}$ & $2.0\times10^{32}$ & $2.0\times10^{22}$ & 32.30 & 22.30 \\
Type III & 10\% & $4\times10^{37}$ & $4.0\times10^{36}$ & $4.0\times10^{48}$ & $4.0\times10^{32}$ & $4.0\times10^{22}$ & 32.60 & 22.60 \\
\bottomrule
\end{tabular}
\end{table}

The three-scenario heatmap makes the relative weight of efficiency progress and energy availability visible at the 2035 horizon. In the \textit{Current} trajectory, achieving even one brain-equivalent of digital compute per human inhabitant on Earth alone requires allocating approximately 50\% of primary energy to cognition; below that, civilisational-scale (Type~I) energy expansion is necessary to clear the parity bar. In the \textit{Better} trajectory, parity is achieved on Earth itself at $f = 5\%$, without any expansion of total power. In the \textit{Optimistic} trajectory, parity is exceeded on Earth at every $f$ examined, including $f = 1\%$, and Type~I-scale energy at $f = 10\%$ buys roughly $10^4$ brain-equivalents per human inhabitant.

The structural lesson is that if compute efficiency continues to improve at the rate implicit in Figure~\ref{fig:trajectory}'s trajectories, the binding constraint on per-capita cognitive abundance shifts decisively from total energy to the political-economy variable $f$. If efficiency stalls at the present 2026 level, by contrast, the binding constraint remains civilisational energy, and Type~I-scale infrastructure becomes the relevant ceiling.

The per-capita parity threshold deserves attention in its own right. ``One brain-equivalent of digital cognition per inhabitant'' sounds abstract, but it is simply the capacity, in compute units, for every human to have access to a personal AI assistant operating continuously at roughly the throughput of their own cortex. At present this kind of access is concentrated at the high end of the income distribution, distributed unevenly through paid subscriptions and professional tools. At the Type~0.73 (Earth 2025) tier in the \textit{Optimistic} trajectory, or at the Type~I tier under any trajectory, the total compute supply is sufficient to make universal access of this kind physically possible. Whether it becomes universally available is a question of political economy and institutional design rather than thermodynamics. Two further patterns stand out. A Type~I civilisation at $f = 1\%$ commands $\sim 10^{11}$ brain-equivalents in the \textit{Current} trajectory, an order of magnitude above global human population; the same tier in the \textit{Optimistic} trajectory commands $\sim 10^{13}$, three orders of magnitude above. A Type~II civilisation operates at $\sim 10^{11}$ brain-equivalents per inhabitant even at modest cognitive allocation in the \textit{Current} trajectory, rising to $\sim 10^{13}$--$10^{14}$ per inhabitant in the \textit{Optimistic} trajectory. The numbers at Type~III are interpretable only as upper bounds on what stellar engineering would notionally permit.

\section{The Trajectory toward the Envelope}\label{sec:trajectory}

The static envelope above describes what a Kardashev civilisation \textit{could} support at given $f$ and $\eta$. The dynamic question is whether and how fast contemporary humanity is moving along the envelope. Three trends are relevant.

\begin{figure}[!htbp]
\centering
\includegraphics[width=0.95\textwidth]{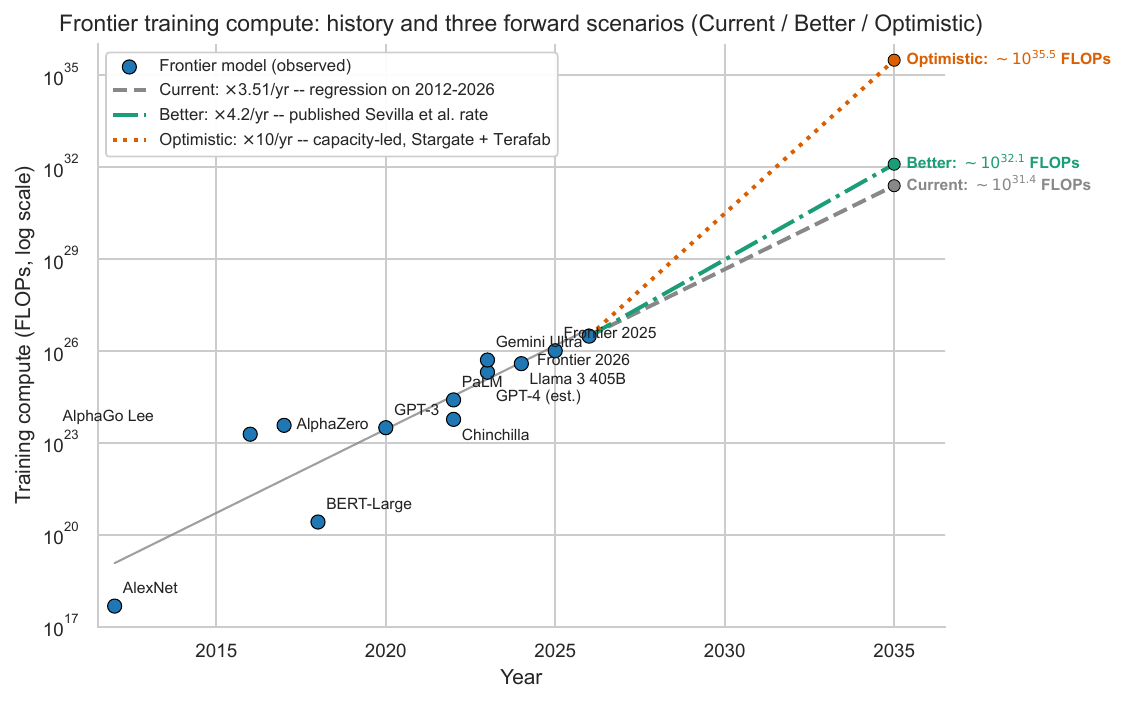}
\caption{\textbf{Frontier training-compute trajectory and three forward scenarios, 2012--2035.} Twelve observed frontier models (blue dots, with the historical regression as the thin grey line). Three forward scenarios projected from the 2026 anchor of $3\times 10^{26}$ FLOPs to 2035: \textit{Current} at $\times 3.51$/yr (grey dashed; regression on 2012--2026), \textit{Better} at $\times 4.2$/yr (green dash-dot; published \citet{Sevilla2022} rate), and \textit{Optimistic} at $\times 10$/yr (orange dotted; capacity-led, Stargate + Terafab buildout). 2035 endpoint values shown bold to the right of each scenario. Sources: \citet{Sevilla2022,EpochAI2024,Maslej2025}; numerical values are order-of-magnitude.}
\label{fig:trajectory}
\end{figure}

First, frontier training compute has grown rapidly. \citet{Sevilla2022} estimate a factor of $\sim$4.2 per year for the deep-learning era. A simple log-linear regression through twelve representative frontier models from AlexNet (2012, $\sim 10^{17}$ FLOPs) to estimated 2026 frontier runs ($\sim 3\times 10^{26}$ FLOPs) yields a slightly more conservative factor of $\sim$3.5 per year (Figure~\ref{fig:trajectory}), corresponding to a doubling time of roughly 6.6 months. For comparison, the classical Moore's-Law doubling time was $\sim$24 months and applied to a much narrower property (transistor count per chip). Frontier-AI compute has, over the past decade, doubled roughly four times faster than transistors did in their long boom, and from a base many orders of magnitude larger. There is no industrial precedent in the modern era for sustained scaling of this kind; the closest analogy may be the early electrification of national economies in the late nineteenth century.

Second, I project three forward scenarios as conditional extrapolations rather than predictions. From a 2026 anchor of $3\times 10^{26}$ FLOPs I extend through 2035 under: (a) a \textit{Current} scenario at $\times 3.51$/yr, the regression slope through the twelve observed frontier models, a slope fragile to point selection (BERT-Large is a known low outlier, and excluding it raises the slope toward Sevilla's value); (b) a \textit{Better} scenario at $\sim$$\times 4.2$/yr, the published \citet{Sevilla2022} estimate for the deep-learning era; and (c) an \textit{Optimistic} capacity-led scenario at $\sim$$\times 10$/yr. The first two are continuations of historical fits; the third is a feasibility benchmark, asked of the data rather than supplied by it. The Stargate consortium's announced 10\,GW target by 2029 \citep{Stargate_2025} and the Terafab venture's terawatt-per-year aspiration \citep{BloombergTerafab2026} make the \textit{Optimistic} rate technically conceivable on the capacity side, but several reports during 2025 indicate Stargate has shifted partly toward leasing rather than de novo build, and Terafab is at announcement rather than groundbreaking stage. The corresponding 2035 frontier-training-run sizes are $\sim 10^{31.4}$, $\sim 10^{32.1}$, and $\sim 10^{35.5}$ FLOPs respectively. The third figure is, as I show next, physically infeasible at fixed efficiency and is therefore a stress test rather than a prediction.

Third, the energetic cost of frontier training is converging on the budgets of medium-sized cities and small countries. A frontier $3\times 10^{26}$-FLOP run at $\eta_{2026} = 10^{12}$ FLOP/J consumes $3\times 10^{14}$~J, or $\sim$83 GWh, comparable to several days of electricity consumption for a small European country such as Estonia ($\sim$21 GWh/day) or about a third of a day for a mid-sized economy such as Belgium ($\sim$220 GWh/day). A single 1.2\,GW data-centre site operating continuously consumes $\sim$10.5 TWh/yr, an order of magnitude more than 100 such training runs.

Decomposing each scenario into its implied energy and efficiency components clarifies feasibility. Compute scales as $C \propto E \cdot \eta$, where $E$ is total energy consumed by training runs and $\eta$ is sustained efficiency. From a 2026 baseline ($E_{2026} \cdot \eta_{2026}$), the 2035 endpoint of each scenario decomposes as
\[
\frac{E_{2035}}{E_{2026}} \cdot \frac{\eta_{2035}}{\eta_{2026}} \;=\; \text{growth factor}.
\]
For the \textit{Current} scenario this growth is $\times 8\!\times\!10^{4}$ (i.e.\ $3.51^{9}$), the \textit{Better} scenario $\times 4\!\times\!10^{5}$ (i.e.\ $4.2^{9}$), and the \textit{Optimistic} scenario $\times 10^{9}$. Distributing this growth between energy expansion and efficiency improvement gives a feasibility envelope. With energy expansion held to a plausible $\times 10^{2}$--$10^{3}$ over nine years (consistent with IEA Base Case data-centre buildout extrapolated to 2035), the \textit{Current} scenario requires $\eta$ to grow by roughly $\times 10^{2}$ (toward $10^{14}$ FLOP/J), the \textit{Better} scenario by roughly $\times 10^{3}$ (toward $10^{15}$ FLOP/J, approaching the lower bound of biological-cortex efficiency), and the \textit{Optimistic} scenario by $\times 10^{6}$--$10^{7}$. The latter places sustained $\eta$ in the range $10^{18}$--$10^{19}$ FLOP/J, which approaches or exceeds the Landauer-FLOP equivalent at 300~K under any reasonable assumption about bit-operations per FLOP. The \textit{Optimistic} scenario at fixed room-temperature irreversible computation is therefore at or near the thermodynamic ceiling and is best read as a stress test, an upper bound on what compute scaling alone can deliver under generous capacity assumptions, achievable only through some combination of reversible computation, sub-room-temperature operation, and very aggressive capacity buildout. The \textit{Current} and \textit{Better} scenarios remain feasible on the 2035 horizon, conditional on substantial but not unprecedented efficiency progress.

The OpenAI--SoftBank--Oracle--MGX \textit{Stargate} consortium, announced in January 2025 with a \$500\,billion commitment over four years, targets 10\,GW of dedicated AI compute capacity by 2029 \citep{Stargate_2025,Stargate_5sites_2025}; the Tesla--SpaceX--xAI \textit{Terafab} initiative, announced by Elon Musk in March 2026 with an initial \$55\,billion (projected \$119\,billion total) investment, targets vertically integrated production of more than one terawatt of AI compute capacity per year \citep{BloombergTerafab2026,CNBCTerafab2026}. The combined committed capital across these and adjacent infrastructure programmes runs into the high hundreds of billions of dollars over a five-year horizon, a scale that exceeds the inflation-adjusted cost of the Manhattan or Apollo Programmes and approaches a major mid-twentieth-century rearmament cycle. Frontier AI compute is now planned and capitalised at a level previously reserved for war or national reconstruction. These commitments make the optimistic scenario more plausible than it would otherwise be; they also concentrate the cognitive substrate in the hands of a small set of vertically integrated firms with national-scale energy footprints. The distributive implications of that concentration are the subject of Section~\ref{sec:hayek}.

\section{Sensitivity and Parameter Uncertainty}\label{sec:sensitivity}

The brain-equivalent count is the product of six quantities once the per-capita normalisation and the data-centre overhead corrections are made explicit:
\[
N_{\text{pc}} \;=\; \frac{f\,P\,\eta\,\phi}{C_{\text{brain}}\,\mathrm{PUE}\,N_{\text{pop}}},
\]
each carrying separate uncertainty. The qualitative structure of the envelope is robust to plausible joint variation, but combined uncertainty is substantial; the envelope as reported in Tables~\ref{tab:envelope}--\ref{tab:big} should be read as carrying $\pm 1.5$ to $2$ orders of magnitude of joint uncertainty when the brain-compute and AI-share parameters are propagated together. Within that range, two qualitative conclusions are robust: Type~I and beyond carry per-capita cognitive surpluses many orders of magnitude above what biology supplies, and Earth 2025 already supplies tens of millions of digital brain-equivalents in raw capacity.

\begin{itemize}
\item \textbf{$P$} is the most empirically pinned; world primary energy is well-measured at $\sim 2.0\times10^{13}$ W, and Type~I, II, III are by definition. Stellar luminosity $L_\odot = 3.828\times 10^{26}$ W is a Solar reference; full-spectrum capture is a physical idealisation \citep{Dyson1960,Sandberg1999}.
\item \textbf{$f$} is the policy variable. Today's data-centre share of $\sim$0.24\% is already an order of magnitude above the level of a decade ago; IEA Base Case projections imply $\sim$0.5--1\% by 2030 \citep{IEA_EnergyAI_2025}. I treat $f \in [10^{-3}, 0.5]$ as the plausible long-run policy range.
\item \textbf{$\eta$} is the technical-progress variable. The two- to three-order-of-magnitude gap between sustained FP64 supercomputers and peak FP4 ML inference suggests that $\eta$ has substantial residual room to grow within current physics. Stagnation at the present sustained-ML value of $\eta_{2026} = 10^{12}$ FLOP/J is unlikely on a multi-decade horizon; further increases toward $10^{14}$--$10^{15}$ FLOP/J are conceivable with neuromorphic, optical, or biological substrates approaching the lower bound of biological-cortex efficiency.
\item \textbf{$\phi$} (AI-share of compute) is poorly constrained. I use $[0.05, 0.30]$ for the global aggregate in 2024, rising to $[0.20, 0.60]$ by 2030. This single parameter introduces approximately one order of magnitude of uncertainty into all aggregate brain-equivalent figures.
\item \textbf{$\mathrm{PUE}$} (data-centre overhead) is empirically narrow; typical hyperscale PUE is $1.1$--$1.5$, with industry averages at $\sim 1.3$. I use $1.3$ throughout. Variation here is sub-leading.
\item \textbf{$C_{\text{brain}}$} is the most contested parameter. The literature spans $10^{15}$ to $10^{18}$ FLOP/s depending on whether one counts synaptic events, dendritic computations, or whole-brain dynamics \citep{Bartol2015,SandbergBostrom2008,LevyCalvert2021}. My $10^{16}$ FLOP/s is mid-range; reported brain-equivalent figures carry $\pm 1$ order of magnitude of irreducible uncertainty from this source alone.
\item \textbf{$N_{\text{pop}}$} is well-measured at $8.1\times 10^{9}$ in 2024 and projected by the UN to $\sim 9.7\times 10^{9}$ by 2050. Tables in this paper use $10^{10}$ as a round number; the implied error is $\sim 25\%$, well below other parameters' uncertainty.
\end{itemize}

\section{Discussion: What the Scale Does and Does Not Show}\label{sec:hayek}

The Scale developed in §§\ref{sec:eta}--\ref{sec:trajectory} is descriptive. It quantifies how much sustained ML-grade computation each Kardashev tier could in principle support, given a defensible 2024--2026 efficiency anchor and a transparent uncertainty range on the biological reference. Three observations follow, none of which the analysis itself fully establishes; they are interpretive remarks calibrated against the numerical content.

First, raw capacity in absolute terms is comfortably in excess of population-weighted parity at planetary energy scales. A Type~I civilisation at $f = 1\%$ commands $10^{26}$ FLOP/s, equivalent to between $10^{9}$ and $10^{11}$ brain-equivalents under the $C_{\text{brain}}$ uncertainty range, and Earth at present energy levels under full allocation already exceeds the same band. The implication is that the physical-capacity question is not the binding constraint at any plausible Kardashev tier, since a civilisation at Type~I or beyond does not run out of cognitive capacity for any realistic per-capita demand. The substantive question is therefore distributional. The Scale does not measure how capacity is allocated across populations, jurisdictions, and uses, and it cannot answer that question.

Second, the trajectory and feasibility analysis of §\ref{sec:trajectory} highlights two distinct regimes. Under stagnation of $\eta$ near the present sustained-ML value, growth in cognitive output requires linear growth in energy, which the IEA forecasts and contemporary infrastructure announcements (Stargate, Terafab) suggest is in fact occurring. Under continued efficiency progress toward $10^{14}$--$10^{15}$ FLOP/J, the lower bound of biological-cortex efficiency, the same compute growth can be delivered with much less marginal energy. Which regime obtains determines whether allocable energy or institutional access becomes the binding margin. The paper does not adjudicate; both regimes are empirically possible on the 2035 horizon.

Third, the empirical pattern of compute concentration in the contemporary AI infrastructure is independent of the Scale but interacts with it. A small number of vertically integrated consortia control the bulk of frontier capacity \citep{Stargate_2025,BloombergTerafab2026,IEA_EnergyAI_2025}. If access to the cognitive surplus implied by Type~I-tier energy is similarly concentrated, the per-capita brain-equivalent figures in Table~\ref{tab:big} describe a population-weighted average that masks an extreme distributional skew. The Scale quantifies what is in principle available; whether and how it is socially allocated is a political question this paper does not engage.

A longer pattern lies behind the calibration. For most of history, cognition was scarce in the way calories were scarce; a person had what their own mind could produce, modestly augmented by tools and books. Mass literacy in the nineteenth century, mass schooling in the twentieth, and now mass-deployed AI in the twenty-first are successive expansions of the cognitive supply curve. Each expansion required physical infrastructure, from printing presses to schools to data centres, and each was accompanied by political contestation over who had access. The Cognitive Kardashev Scale is scaffolding for the question that recurs at each expansion, namely what binds once the supply of cognition no longer does, and what kind of society gets built around the new constraint. The arithmetic does not answer that question. It does suggest the question has arrived.

\section{Conclusion}\label{sec:conclusion}

Contemporary humanity sits at $K \approx 0.73$ on Sagan's logarithmic refinement of the original Kardashev energy scale. The Cognitive Kardashev Scale developed in this paper runs in parallel with that energetic scale, quantifying what each Kardashev tier could in principle support in sustained ML-grade compute, given current and near-future hardware efficiencies and a transparent uncertainty range on the biological reference. The exercise is calibration, not prediction.

Three things follow from the calibration. (i) Raising $f$ from 1\% to 10\% within a Kardashev tier buys a factor of 10 in cognitive capacity. (ii) Moving from Type~0.73 to Type~I at fixed $f$ buys a factor of $\sim$500 (between two and three orders of magnitude). (iii) Frontier training compute, growing at the historical regression rate, will exceed reasonable global energy budgets by the mid-2030s in the absence of substantial efficiency progress; the \textit{Optimistic} ($\times 10$/yr) trajectory is feasible only if $\eta$ improves by several orders of magnitude as well, ultimately approaching the biological-cortex efficiency at the lower end of the brain-compute range. The \textit{Current} ($\times 3.5$/yr) trajectory is feasible under the more modest efficiency progress empirically observed across the Blackwell-to-Rubin transition and prior generations.

Whether the binding constraint on cognitive abundance is allocable energy or computational efficiency therefore depends on which of these regimes is realised. The paper does not adjudicate. The analysis above answers the abstract question of physical possibility in the affirmative. What remain open are the engineering trajectory of $\eta$, the political and ecological allocation of $f$, and the institutional distribution of access to compute. Each of these is the subject of a different literature, and the calibration developed here is offered as a quantitative input to those literatures rather than as a substitute for them.

The exercise is offered in the spirit of \citet{Dyson1960}'s back-of-envelope argument that stellar-scale energy capture is consistent with known physics, or of Drake's equation as a scaffolding for thinking about civilisational questions whose answers cannot yet be computed. What such calibrations do is convert vague intuitions about ``vast'' or ``unimaginable'' scales into specific numbers that can be argued with and revised as evidence accumulates. The Cognitive Kardashev Scale makes specific the intuition that the long-run constraint on civilisational thought is energy and politics, not silicon. Whether that intuition survives the next decade of evidence is for that decade to reveal. The calibration will be the easier to update because its parameters, ranges, and conditional structure are written down explicitly.

\bibliographystyle{plainnat}
\bibliography{refs}

\end{document}